\documentclass{article}
\usepackage{spconf,amsmath,graphicx}
\usepackage{pgfplots}
\pgfplotsset{compat=1.5}
\usepackage{tikz}
\usepackage{subfig}
\usepackage{bm}
\usepackage{url}
\usepackage{balance}
\usepackage{hhline}
\usepackage{multirow}
\usepackage{booktabs}
\usepackage{tabularx}
\usepackage{hyperref}
\usetikzlibrary{shapes.geometric}
\usetikzlibrary{positioning}
\usetikzlibrary{external}
\AtBeginShipout{\AtBeginShipoutUpperLeft{%
		\put(\dimexpr\paperwidth-20cm\relax,-0.5cm){\parbox[t]{19cm}{\copyright 2023 IEEE. Personal use of this material is permitted. Permission from IEEE must be obtained for all other uses, in any current or future media, including reprinting/republishing this material for advertising or promotional purposes, creating new collective works, for resale or redistribution to servers or lists, or reuse of any copyrighted component of this work in other works. \url{https://doi.org/10.1109/ICASSP48485.2024.10447125}}}%

}}

\newlength\figureheight
\newlength\figurewidth
\def\Figu#1{{ Figure \ref{#1}}}

\def\Tabl#1{{ Table \ref{#1}}}

\title{Improved Screen Content Coding in VVC Using Soft Context Formation}
\name{Hannah Och$^{\star}$ \qquad  Shabhrish Reddy Uddehal$^{\dagger \star}$ \qquad Tilo Strutz$^{\dagger}$ \qquad Andr\'{e} Kaup$^{\star}$\thanks{
		This work has been funded by the Deutsche Forschungsgemeinschaft (DFG, German Research Foundation) - project ID 438221930.
}}
\address{$^{\star}$Multimedia Communications and Signal Processing\\Friedrich-Alexander Universität Erlangen-Nürnberg (FAU), Erlangen, Germany \\
	$^{\dagger}$Department of Electrical Engineering and Computer Science\\
	Coburg University of Applied Sciences and Arts, Coburg, Germany}
\begin{document}
\ninept
\maketitle
\begin{abstract}
Screen content images typically contain a mix of natural and synthetic image parts. Synthetic sections usually are comprised of uniformly colored areas and repeating colors and patterns. In the VVC standard, these properties are exploited using Intra Block Copy and Palette Mode. In this paper, we show that pixel-wise lossless coding can outperform lossy VVC coding in such areas. We propose an enhanced VVC coding approach for screen content images using the principle of soft context formation. First, the image is separated into two layers in a block-wise manner using a learning-based method with four block features. Synthetic image parts are coded losslessly using soft context formation, the rest with VVC. We modify the available soft context formation coder to incorporate information gained by the decoded VVC layer for improved coding efficiency. Using this approach, we achieve Bjontegaard-Delta-rate gains of 4.98\% on the evaluated data sets compared to VVC.
\end{abstract}
\section{Introduction}
\label{sec:intro}
With an increased use of applications such as online video conferences, e-learning and desktop sharing the efficient compression of screen content images (SCIs) gains more importance. SCIs generally consist of synthetic parts, e.g. text, buttons or computer graphics as well as natural image parts, e.g. camera-captured scenes or realistic computer animations. Synthetic image parts differ in their characteristics from natural image scenes. They often contain large uniformly colored areas, sharp contrasts, fine structures, few unique colors, repeating patterns and are generally noise-free.
Due to these differences in image statistics, it is very challenging to compress SCIs. A single compression technique typically cannot effectively compress both types of content. State-of-the-art video coding standards such as \textit{High Efficiency Video Coding} (HEVC) \cite{Sul12,Xu16} or its successor \textit{Versatile Video Coding} (VVC) \cite{Bro21,VVCSCC} have thus introduced multiple coding tools to aid in the compression of synthetic areas. The most prominent tools are palette mode \cite{Wei16}, which exploits the limited number of colors in screen content, and intra block copy (IBC) \cite{Xu16b,Xu19}, which makes use of similar repeated image patches.

Going one step further, many compression methods are based on segmenting an image into separate regions where completely different coding methods are used depending on the type of content. In the ITU-T mixed raster content (MRC) standard \cite{Que98} compound images are separated into components with different image characteristics, namely foreground, background, and a mask plane. Each of these layers is compressed using a different compression algorithm \cite{Kur12}. A hybrid compression method, where synthetic regions are encoded using run length coding and natural regions via JPEG is described in \cite{Mog99}. The united coding method \cite{Wan14} is a block-wise mixed lossy and lossless codec, where for each block first an optimal lossless coder is chosen from a range of available codecs. Then, depending on a rate-distortion optimization criterion, it may be replaced by a lossy coding method. In \cite{Tan22} text semantic-aware coding of SCIs is proposed, where SCIs are separated into textual information and background. The background is encoded using HEVC, while the textual information is encoded separately from the image and restored in the decoder based on size, font, color and positional information. For lossless coding of SCIs, a block-based segmentation method is presented in \cite{Udd23}, where an image is segmented in natural and synthetic blocks, so that their statistics can be learned separately.

For synthetic content, context-based arithmetic coders have been shown to work remarkably well. One such method is the Soft Context Formation (SCF) coder \cite{Och21}, which outperforms VVC and HEVC in the lossless case for synthetic images with few colors and many repeating image structures and even surpasses lossy VVC compression for some cases, as we will see in Section \ref{method}. Thus, we propose to enhance VVC by first segmenting an image into two layers. One layer contains synthetic content which is coded losslessly using SCF, while the rest of the image is coded using VVC. This paper shows that using a simple learned CTU-wise classification method, we can achieve BD-rate gains when compared to the original VVC for SCIs. The paper is organized as follows: First, the key components of the SCF coder are shortly reviewed. Second, the proposed approach is explained in detail. Finally, the results are evaluated using multiple screen content data sets and compared to the VTM 17.2.

\section{Review of Soft Context Formation Coder}
\label{sec:SCF}
\begin{figure}[t]
	\begin{center}
	\hfil\tikzstyle{block} = [draw, fill=white, rectangle, minimum height=1.5em, minimum width=5em, align=center]
\tikzstyle{decision} = [draw, fill=white,diamond, aspect=2,minimum height=2em, minimum width=8em,align=center, inner sep=0pt, outer sep=0pt]
\tikzstyle{input} = [coordinate]
\tikzstyle{output} = [coordinate]
\tikzstyle{pinstyle} = [pin edge={to-,thin,black}]
\tikzstyle{pinstyle2} = [pin edge={-to,thin,black}]

\tikzset{font=\scriptsize}

\begin{tikzpicture}
	\noindent
	\node [decision, pin={[pinstyle]above, pin distance=1.5ex:Get current pixel $X$}, node distance=1cm] (stage1decision){$X$ can be\\ coded in Stage 1};
	\node [decision,below of=stage1decision, node distance=1.5cm] (stage2decision){$X$ is in\\ color palette};
	\node [block, right of=stage1decision, node distance=2.8cm] (stage1){Code $X$ in Stage 1};
	\node [block, right of=stage2decision, node distance=2.8cm] (stage2){Code $X$ in Stage 2};
	\node [block, below of=stage2, node distance=1.25cm] (stage3){Code $X$ in Stage 3};
	\node [block, below right=0.25cm and -0.9cm of stage3,pin={[pinstyle2]below, pin distance=1.5ex:Goto next pixel}] (update){Update histograms\\and pattern lists};
	
	\draw  [->] (stage1decision) -- node [name=Yes1, midway, above] {Yes} (stage1);
	\draw  [->] (stage2decision) -- node [name=Yes2, midway, above] {Yes} (stage2);
	\draw  [->] (stage1decision) -- node [name=No1, midway, left] {No} (stage2decision);
	\draw [->] (stage2decision) |- node [name=No2, near start, left] {No} (stage3);
	\draw [->] (stage1) -| (update);
	\draw [->] (stage2) -| (update);
	\draw [->] (stage3) -| (update);
\end{tikzpicture}
	\vskip -0.8em
	\caption{\label{block_diagram}Block diagram of the SCF method with its three stages.}
	\end{center}
	\vskip -0.9em
\end{figure}
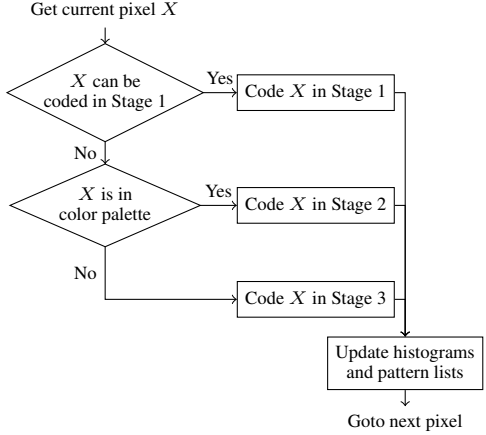

The SCF coder \cite{Och21,Str19} is a pixel-wise lossless entropy coder developed for SCIs. The main processing stages can be seen in \Figu{block_diagram}.
The image is processed in a raster-scan order. In Stage 1 of the SCF coder, for each pixel a probability distribution of the current RGB value is estimated based on the already encoded pixels. Therefore, as can be seen in \Figu{fig_template_matching}, the neighboring pixels in pattern $P = \{A,B,C,D,E,F\}$ are utilized as a context. During the coding process, for each context the associated colors of pixel $X$ are saved in a histogram. For the current pixel $X$, the color histograms of similar patterns are combined and used for the generation of a probability distribution, which then enables the coding of the RGB color value  using an arithmetic coder. Given $n(c|P)$ as the count of color $c$ in the combined histogram of similar patterns of $P$, and $N_P$ the total number of counts in the histogram, the probability of color $c$ given pattern $P$ is estimated as 	%
\begin{equation}
	p(c|P) \approx \frac{n(c|P)}{N_P}.
\end{equation}
In case the current pixel color is part of the combined histogram, the pixel value will be encoded using an arithmetic encoder, which roughly needs the self-information 	%
\begin{equation}
	I(c|P)=-\log_2{p(c|P)}\; \text{[bit]}
\end{equation}
to encode the entire RGB color. However, if the generated histogram for the current pattern $P$ does not yet contain the color $c$, e.g., when a color happens for the first time in combination with pattern $P$, an escape symbol is sent and the second stage of the SCF coder is initiated. In Stage 2, a probability distribution is generated based on a global color palette. The color palette is updated after every pixel coded in Stage 2 or 3 and contains all already seen colors and their occurrences except for pixels coded in Stage 1. If a color has not occurred before, this is signaled via an escape symbol and the color will be encoded in Stage 3 using residual coding: After predicting the pixel color in a channel-wise manner using an enhanced median adaptive predictor (cMAP) \cite{Str19}, the prediction error of each channel is encoded arithmetically based on a histogram of previously encoded prediction errors.
\section{Proposed method}
\label{method}
Inspired by the high gain of the SCF coder against VVC in the lossless case for certain types of images, we have investigated the coding performance of SCF compared to lossy VVC coding. We compare the bit rates needed using SCF as well as VVC with different quantization parameters (QPs) for small image areas. To this end, 8602 blocks  of size $128\times128$ are taken from 236 SCIs from own resources. Each block is separately compressed using SCF and the VVC reference codec VTM 17.2 for QP 22, 27, 32 and 37. For the VTM all-intra configuration for RGB data and camera-captured content according to \cite{JVET_T2013} is used, i.e. the VTM is applied in the RGB 4:4:4 setting with all screen content specific coding tools, such as palette mode, IBC, transform skip residual coding \cite{VVCTSRC} or block-based differential pulse-code modulation \cite{VVCBDPCM}, enabled. Even with these coding tools enabled in the VVC, the SCF coder has a better lossless compression for 49.5\%, 43.7\%, 38.9\% and 34.1\% of blocks than the lossy VTM with QP 22, 27, 32 and 37, respectively. This means that, for QP 22, almost half of the blocks need less bit rate when encoded with SCF. Even for QP 37 still more than a third of the blocks profits from lossless SCF coding in terms of bit rate without introducing distortion. In the following, we describe how we utilize this information by first segmenting an SCI into two layers and then coding the layers separately with VVC and SCF.

\subsection{Block-wise segmentation}
\label{segmentation}
\begin{figure}[t]
	\hfil\includegraphics[scale=0.16]{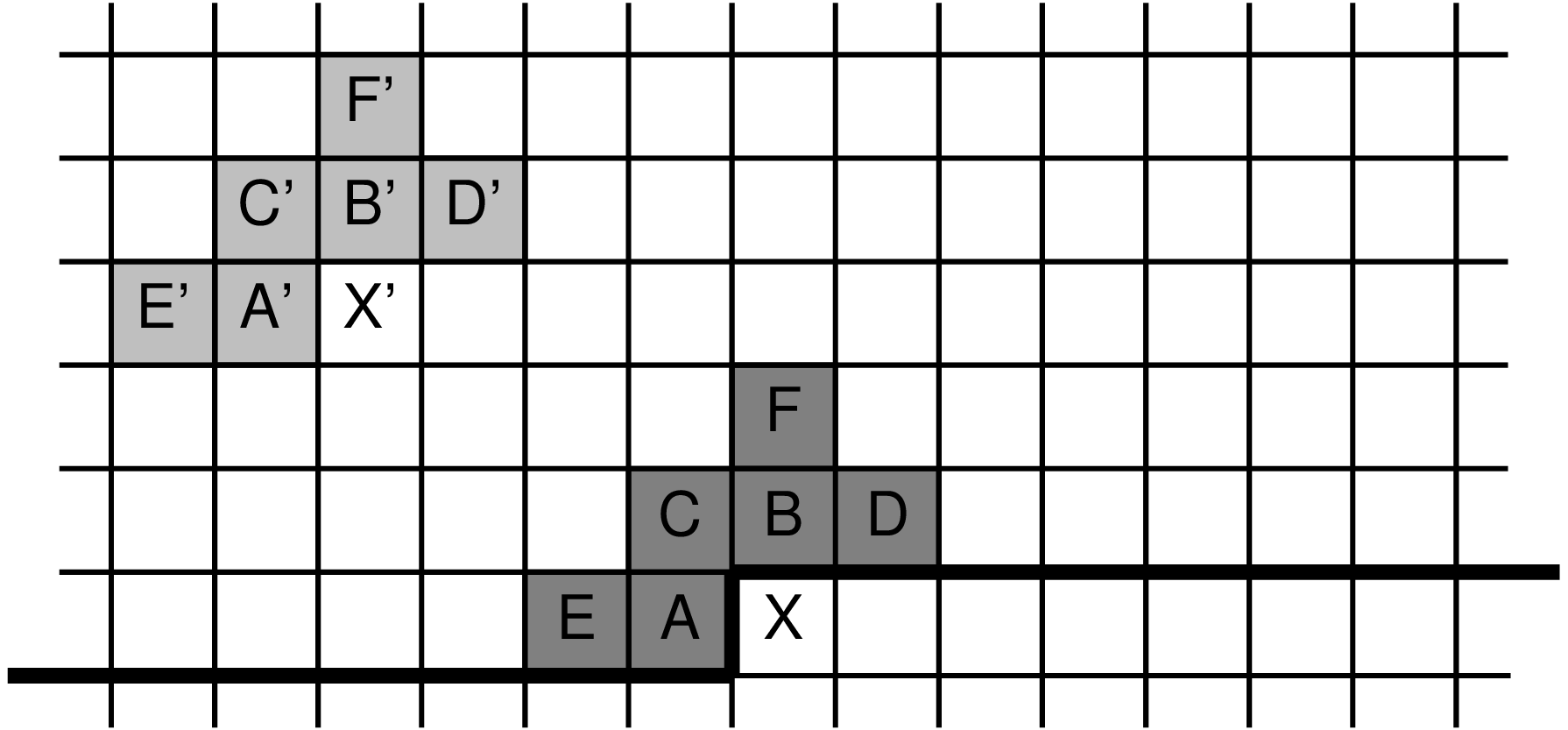}
	\vskip -0.8em
	\caption{\label{fig_template_matching}Context pattern: If the values in the template $\{A,B, \dots, F\}$ are similar to $\{A',B', \dots, F'\}$, then the current value at position $X$ is likely to be similar to the value at $X'$.}
	\vskip -0.9em
\end{figure}

The first step of the proposed method is a segmentation of the image into two layers. The SCF layer contains synthetic image sections and is encoded using SCF, while the VVC layer contains the remaining image parts and is compressed using the VVC. It is important to note that the goal of the segmentation is not a perfect segmentation into natural and synthetic image parts. 

Since the SCF coder is lossless, its usage for some image parts will reduce image distortion in general. Hence, to simplify the segmentation, we can focus on rate-optimization, which directly causes an improvement of rate-distortion. For a useful segmentation, multiple factors have to be taken into account:
First, the segmentation mask should be easy to transmit without a large overhead. Second, the segmentation should not affect the VVC coding performance significantly. Both of these points lead us to the decision to segment block-wise, using blocks of the same size as the CTUs in VVC, in this case $128\times128$, which can easily be transmitted using 1 bit per CTU. 
As a last point to consider, syntheticness does not directly translate to a good SCF coding performance. E.g., a constant vertical and horizontal color gradient can be synthetically generated, but is very difficult to compress using the SCF coder, since it has almost no repeating patterns and also many colors. Consequently, a classification method specifically adapted for SCF has to be applied instead of simply using existing natural/synthetic segmentation methods.

It is difficult to estimate the rate needed for a CTU when using the SCF coder without actually encoding the block, and it is computationally expensive for the VVC coder. As such, we propose a segmentation algorithm, where we estimate whether the bit rate needed to encode a CTU using the SCF coder is less than the bit rate needed with VVC. To this end, we train a k-nearest neighbor classifier based on four features, which classifies whether a CTU fulfills the condition $r_{\mathrm{SCF}} < r_{\mathrm{VVC},i}$ with $r_{\mathrm{SCF}}$ as the SCF bit rate and $r_{\mathrm{VVC},i}$ with $i \in \{22,27,32,37\}$ the QP-dependent VVC bit rate for one CTU. Based on the classification result, the CTU is assigned to the SCF or VVC layer accordingly. 

For the SCF coder, the main features influencing its coding performance are number of colors, number of unique patterns as well as the entropy of colors and  the entropy of a pixel color given its context. These four features are thus taken as predictors for the classifier. In this case, the counted patterns are simplified to incorporate only the direct neighbors plus the current pixel: $P_X' =\{A,B,X\}$. The entropy of colors is calculated only on pixels predicted to be encoded in Stage 2 and Stage 3. Stage 1 pixels should be excluded to simulate the color palette as used in Stage 2 (see Section \ref{sec:SCF}). To this end, all pixels for which a specific pattern $P_X'$ happens for the first time are estimated to have to be encoded in Stage 2 or 3. Furthermore, the entropy of a pixel color given its context  $H(X|P' =\{A,B\})$ is also calculated using the simplified pattern and simulates the bit rate needed for encoding a pixel in Stage 1. To enable the use of the classifier for different block sizes such as cut-off blocks at image edges, the number of colors and number of patterns are normalized by the number of pixels per block. 

\begin{figure}[!tbp]
	\centering
	\subfloat[Original image.]{\includegraphics[width=0.2\textwidth]{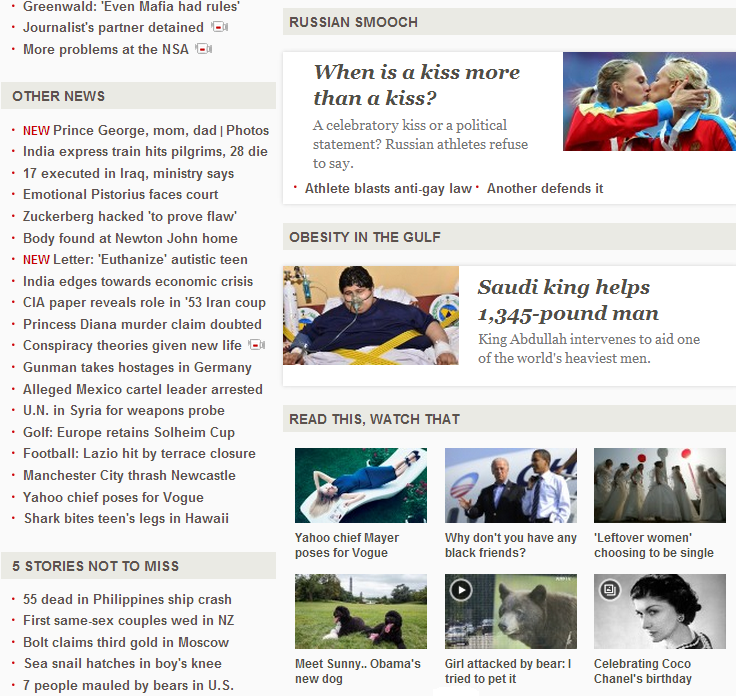}\label{fig:f1}}
	\hfill
	\subfloat[Segmented image.]{\includegraphics[width=0.2\textwidth]{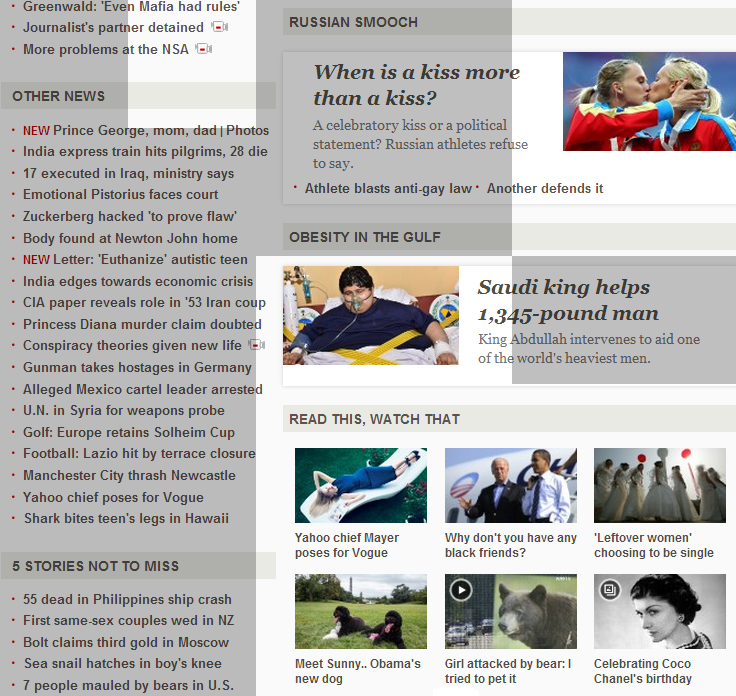}\label{fig:f2}}
	\vskip -0.8em
	\caption{Image segmented block-wise into separate layers coded by the VVC and the SCF coder for QP 22. Grayed out areas indicate pixels to be encoded in the SCF layer.}
	\label{segmentationIm}
	\vskip -0.9em
\end{figure}

Since $r_{\mathrm{VVC},i}$ is QP-dependent, we need to train one classifier per QP. We utilize the 8602 blocks  of size $128\times128$ mentioned above and 10-fold cross-validation. The trained classifiers reach a validation accuracy of $90.6\%$, $91.8\%$, $92.4\%$ and $92.6\%$ for QP 22, 27, 32 and 37, respectively. An example image segmented in this manner using the classifier trained for QP 22 can be seen in \Figu{segmentationIm}. Most completely textual blocks are contained in the SCF layer with the exception of one block where small image icons are present. The natural images are contained in the VVC layer, as expected.

\subsection{Coding Process}
\label{coding}
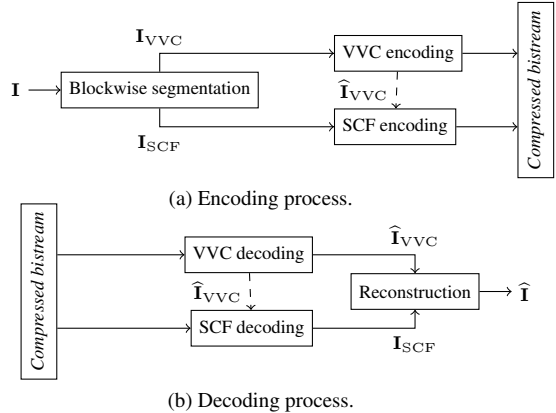
\begin{figure}[t]
	\centering
	\subfloat[Encoding process.]{\tikzstyle{block} = [draw, fill=white, rectangle, minimum height=1.5em, minimum width=5em, align=center]
\tikzstyle{decision} = [draw, fill=white,diamond, aspect=2,minimum height=2em, minimum width=8em,align=center, inner sep=0pt, outer sep=0pt]
\tikzstyle{input} = [coordinate]
\tikzstyle{output} = [coordinate]
\tikzstyle{pinstyle} = [pin edge={to-,thin,black}]
\tikzstyle{pinstyle2} = [pin edge={-to,thin,black}]

\tikzset{font=\scriptsize}

\begin{tikzpicture}
	\noindent
    \node [block, pin={[pinstyle]left:$\mathbf{I}$}, node distance=1.0cm] (segmentation){Blockwise segmentation};
		\node [block, above right=0.0cm and 1.0cm of segmentation] (VVC){VVC encoding};
		\node [block, below right=0.0cm and 1.0cm of segmentation] (SCF){SCF encoding};
		\node [block, right of=segmentation, rotate=90, node distance=5cm] (bitstream){\textit{Compressed bistream}};
		\draw  [->] (segmentation) |- node [name=Yes1, above] {$\mathbf{I}_{\mathrm{VVC}}$} (VVC);
		\draw  [->] (segmentation) |- node [name=Yes1, below] {$\mathbf{I}_{\mathrm{SCF}}$} (SCF);
		\draw  [->, dashed] (VVC) -- node [name=No1, midway, left] {$\widehat{\mathbf{I}}_{\mathrm{VVC}}$} (SCF);
		\draw  [->] (VVC) -- (VVC-|bitstream.north);
		\draw  [->] (SCF) -- (SCF-|bitstream.north);

\end{tikzpicture}} \\
	\vskip -0.5em
	\subfloat[Decoding process.]{\tikzstyle{block} = [draw, fill=white, rectangle, minimum height=1.5em, minimum width=5em, align=center]
\tikzstyle{decision} = [draw, fill=white,diamond, aspect=2,minimum height=2em, minimum width=8em,align=center, inner sep=0pt, outer sep=0pt]
\tikzstyle{input} = [coordinate]
\tikzstyle{output} = [coordinate]
\tikzstyle{pinstyle} = [pin edge={to-,thin,black}]
\tikzstyle{pinstyle2} = [pin edge={-to,thin,black}]

\tikzset{font=\scriptsize}

\begin{tikzpicture}
	\noindent
    \node [block, pin={[pinstyle2]right:$\mathbf{\widehat{I}}$}, node distance=5cm] (reco){Reconstruction};
		\node [block, left of=reco,rotate=90, node distance=5cm] (bitstream){\textit{Compressed bistream}};

		\node [block, above left=0.0cm and 0.5 cm of reco] (VVC){VVC decoding};
		\node [block, below left=0.0cm and 0.5 cm of reco] (SCF){SCF decoding};
		\draw  [->] (VVC-|bitstream.south) -- (VVC);
		\draw  [->] (SCF-|bitstream.south) -- (SCF);
		\draw  [->] (SCF) -| node [name=Yes1, midway,below] {$\mathbf{I}_{\mathrm{SCF}}$} (reco);
		\draw  [->, dashed] (VVC) -- node [name=No1, midway, left] {$\mathbf{\widehat{I}}_{\mathrm{VVC}}$} (SCF);
		\draw  [->] (VVC) -| node [name=Yes1, midway, above] {$\mathbf{\widehat{I}}_{\mathrm{VVC}}$} (reco);
\end{tikzpicture}} \\
	\vskip -0.8em
	\caption{\label{codingProcess}Block diagrams of the encoding and decoding process of the proposed method.}
	\vskip -0.9em
\end{figure}

The proposed modification to the VVC coding method can be described as follows: First, in a pre-processing step, the image is segmented into a VVC layer and an SCF layer as described in Section \ref{segmentation}. Afterwards, to reduce unnecessary overhead, a mode flag signals whether the current image contains only SCF mode blocks, only VVC mode blocks, or both. Additionally, height and width of the image are signaled. If both VVC and SCF modes are contained in the image, a flag signals the label for each CTU. This flag can be transmitted using 1 bit per block and is thus relatively cheap. The VVC layer is encoded in RGB 4:4:4 format using the VTM 17.2 reference codec with all intra configuration and the configurations for non-4:2:0 computer generated content, as described in \cite{JVET_T2013}. All SCF CTUs are set to black. The rest of the image is encoded using the SCF coder. Therefore, the SCF coder implementation is adapted to skip all pixels that are not included in the SCF layer. The basic processing blocks are visualized in \Figu{codingProcess}.
Furthermore, the base SCF coder from \cite{Och21} is modified to utilize information gained by the available decoded VVC layer. To this end, the SCF layer image is filled with the decoded VVC layer $\widehat{\mathbf{I}}_{\mathrm{VVC}}$ at all positions not labeled SCF. This makes a difference for the SCF coder, since the patterns (see \Figu{fig_template_matching}) at CTU edges may include pixels from the VVC layer. Additionally, if colors at CTU edges are new, the cMAP prediction utilizes pixels from the previous block. Moreover, a color palette is generated from the VVC layer and sorted in descending order of number of occurrences. The SCF encoder calculates the percentage of colors that intersect between the SCF layer image and the decoded VVC palette for the $\lfloor 1/{2^b} \rfloor$ most occurring colors in the VVC palette for $b \in {0, ..., 6}$. The value $b$ for the highest percentage is encoded, where $b=7$ is encoded if the highest percentage is smaller than 0.9. This cut off has been empirically chosen to prevent incorporation of too many unneeded colors that would disturb Stage 2 of the SCF coder. The SCF palette is initialized with the partial VVC palette and their color counts are normalized with the block width. Consequently, $b$ can be encoded using only three bits and still allows for usage of the entire palette up to only $1/64$th or even nothing at all of the pre-decoded VVC palette.

The decoding process follows the same pipeline: First, the mode flag is decoded, followed by width and height and in the mixed case also the block labels. Then, the VVC layer is decoded, and used for initialization of the palette as well as the pixels at the edge of the SCF layer. The SCF coder decodes the SCF layer and reconstructs the image from both decoded layers.

\section{Evaluation}
For evaluation, the proposed SCF-enhanced VVC codec is compared with the VTM 17.2. As test data, we use multiple available RGB SCI data sets with a total of 173 images. All test images have an uncompressed bit rate of 24 bpp.  
Test set HEVC-CTC is comprised of single frames taken from HEVC CTC test sequences listed as \textit{Test set 2} in \cite{Str19} from the common test conditions for screen content coding \cite{HEVC_SCC_CTC} and screen content range extension experiments \cite{HEVC_RCE3}. Additionally, the well-known SCI data sets SIQAD \cite{Hua15} and SCID \cite{SCID,Zha17} containing 20 and 40 images, respectively, are evaluated. Finally, the `text' and `mixed' stimuli (SC-Text and SC-Mixed) of \cite{She14} contain screenshots of web pages with a resolution of $1360\times 768$ consisting of textual content in combination with natural images parts. For the test sets, the bit rate is averaged over all images. The RGB-PSNR is averaged as a quality metric. Additionally, the RGB-SSIM and the GFM \cite{GFM} score
are calculated to simulate subjective quality. We use GFM in addition to the well-known SSIM, since it is a quality metric specifically developed for screen content and is able to capture the characteristics of SCIs better. Both GFM and SSIM are averaged over all images after transformation into the log-domain via $s_{\mathrm{log}} = -\log_{10}(1-s)$, where $s$ is the quality score of either SSIM or GFM and $s_{\mathrm{log}}$ is the transformed score.

One image is classified entirely as an SCF image for one QP and thus is encoded without any loss. This image (\textit{`SCI34'}) from the SCID data set is removed from all averaged bit rates and quality scores, since it impedes meaningful Bjontegaard-Delta calculation.

\Tabl{pixSCFperQP} shows the percentage of pixels per QP and per data set which are coded using SCF. As can be seen, the percentage varies over the data sets, depending on the image content, e.g., roughly 20\% for SIQAD up to more than half in test set HEVC-CTC and SC-Text. Additionally, the percentage of SCF pixels decreases with rising QP. On average, 39.2\% of the pixels are labeled as SCF pixels for QP 22, while roughly a fourth of the pixels for QP 37 are part of the SCF layer. This is due to the fact, that with higher QP, VVC needs less bit rate per block and thus, it is less likely that the lossless SCF coder is less expensive in terms of bit rate.

\begin{figure}[t]
	\centering
	\hfil
	\setlength\figurewidth{0.8\columnwidth}
	\setlength\figureheight{.45\columnwidth}
	\centering
\definecolor{mycolor1}{rgb}{0.00000,0.44700,0.74100}%
\definecolor{mycolor2}{rgb}{0.85000,0.32500,0.09800}%
\begin{tikzpicture}

\begin{axis}[%
width=0.951\figurewidth,
height=\figureheight,
at={(0\figurewidth,0\figureheight)},
scale only axis,
xmin=0.3,
xmax=0.9,
xlabel style={font=\color{white!15!black}},
xlabel={Rate in bpp},
ymin=36,
ymax=52,
ylabel style={font=\color{white!15!black}},
ylabel={PSNR in dB},
axis background/.style={fill=white},
xmajorgrids,
ymajorgrids,
legend style={at={(0.97,0.03)}, anchor=south east, legend cell align=left, align=left, draw=white!15!black}
]
\addplot [color=mycolor1, dashed, line width=0.8pt, mark=square*, mark options={solid, fill=mycolor1, mycolor1}]
  table[row sep=crcr]{%
0.861423087960731	50.7345362332588\\
0.630532262796584	46.5792712986744\\
0.451142273493057	42.3772485736322\\
0.315254975998584	37.9859471533329\\
};
\addlegendentry{VTM 17.2}

\addplot [color=mycolor2, line width=0.8pt, mark=*, mark options={solid, fill=mycolor2, mycolor2}]
  table[row sep=crcr]{%
0.844406688368196	51.059295914094\\
0.61587964101123	46.9529628818153\\
0.443274808630731	42.8578755819313\\
0.315514416888321	38.459023125131\\
};
\addlegendentry{Proposed}

\end{axis}
\end{tikzpicture}%
	\vskip -0.8em
	\caption{\label{Tall_RD}Rate-distortion curve averaged over all test sets with respect to the PSNR for the proposed method and the original VTM 17.2.}
	\vskip -0.9em
\end{figure}
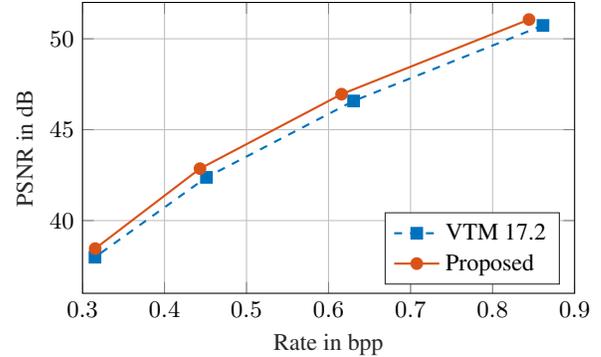

In \Figu{Tall_RD}, the rate-distortion curves with respect to the PSNR, averaged over all test sets, are depicted. The proposed method outperforms the VTM 17.2 for all QPs. The gains are slightly higher for lower QP values. This has two reasons: First, since more pixels are classified as SCF for lower QPs, the proposed method has more opportunity to save bit rate. Second, as the SCF coder is lossless, the bit rate is constant over all QPs. Consequently, the potential bit rate gain increases with decreasing QP when compared to the VVC. Furthermore, the overhead of coding an image using VVC plus SCF will have a higher impact with lower bit rate. To quantify the effectiveness of the proposed method numerically, the Bjontegaard-Delta rates with respect to the VTM 17.2 are calculated using Akima interpolation, as suggested in \cite{Her22}. The results are noted in \Tabl{BDrates}. For all data sets, BD-rate gains can be observed with respect to the PSNR, SSIM and GFM. For the test sets a minimum of 1.48\% and a maximum of 8.19\% PSNR-based BD-rate gains are achieved. Due to the diverse BD-rate results over the data sets, we can conclude that the effectiveness of the proposed method is highly dependent on the image content. Images with a high amount of synthetic content that is clearly separated from the pictorial content, such as the web pages from SC-Text \cite{She14}, profit the most.
On average, the proposed method achieves improvements of 4.98\%, 5.34\%, and 3.96\% BD-rate gains for PSNR, SSIM and GFM, respectively. 
\begin{table}[t]
	\centering
	\hfil

\begin{tabular}{|l||c|c|c|c|} 
\hline
\multicolumn{5}{|c|}{Percentage of pixels in the SCF layer}  \\ 
Data set                & QP 22              & QP 27 			 & QP 32 & QP 37                  \\ \hline
% \hhline{|=::====|}
%Test set 1             & 59.7\% & 52.1\% & 41.7\% & 35.2\%                   \\
%Test set 2             & 51.5\% & 48.9\% & 47.4\% & 43.1\%                     \\
%Test set 3             & 44.2\% & 40.1\% & 36.8\% & 32.5\%                  \\
%Test set 4             & 21.9\% & 19.5\% & 17.8\% & 13.8\%                     \\
{HEVC-CTC \cite{HEVC_SCC_CTC,HEVC_RCE3}}   	& 50.8\% & 47.0\% & 44.3\% & 41.1\%                  \\
{SIQAD \cite{Hua15}}             	& 20.7\% & 19.0\% & 17.9\% & 13.5\%                  \\
{SCID \cite{SCID}}             		& 22.5\% & 20.3\% & 18.0\% & 13.4\%                  \\
{SC-Text \cite{She14}}            & 52.2\% & 49.0\% & 45.1\% & 34.8\%                   \\
{SC-Mixed \cite{She14}}           & 37.3\% & 34.8\% & 31.0\% & 24.6\%                   \\
\hhline{|=::====|}
%All										 & 42.4\% & 38.3\% & 34.4\% & 29.9\%                \\
All															  & 39.2\% & 36.4\% & 33.2\% & 26.4\%                \\

\hline
\end{tabular}
	\vskip -0.8em
	\caption{\label{pixSCFperQP}Percentage of pixels classified as SCF layer pixels.}
	\vskip -0.9em
\end{table}

\begin{table}[t]
	\centering
	\hfil
	{

\begin{tabular}{|l|c||c|c|c|} 
\hline
\multicolumn{5}{|c|}{BD-rate in \% (Reference: VTM 17.2)}  \\ 
Data set                			& \# images             & PSNR & SSIM & GFM                  \\ 
\hline	
{HEVC-CTC \cite{HEVC_SCC_CTC, HEVC_RCE3}} & {14}    & -8.12	& -8.97& 	-6.99 \\  
{SIQAD \cite{Hua15}} & {20}    & -5.19	& -4.09	&-4.26\\  
{SCID \cite{SCID}} & {39}    & -2.67 &	-3.02&	-2.04 \\ 							 
{SC-Text \cite{She14}} & {50}    & -8.19&	-8.36	&-6.97 \\				
{SC-Mixed \cite{She14}} & {49}     & -1.48 &	-2.68	&-0.76 \\ \hline												 
\hhline{|==::===|}
{All}      	& {173}				     & -4.98  & -5.34  & -3.96 \\																
\hline
\end{tabular}
	}
	\vskip -0.8em
	\caption{\label{BDrates}BD-rate results in \% for all data sets with respect to VTM 17.2, negative numbers denote rate savings.}
	\vskip -0.9em
\end{table}
When processing an image block-wise with different coding methods, especially if some blocks are coded lossless, there may be a noticeable visual quality difference. In the evaluated test sets, for QPs 22 to 32, these differences are negligible. For higher QPs, rarely a difference is visible between CTUs coded in different layers. This is generally only the case, if a neighboring CTU with similar content to an SCF CTU is coded by VVC and sharp edges or slight noise artifacts are smoothed.

In terms of computational complexity, the proposed method actually requires less encoding time, due to the reduced rate-distortion optimization search in the VVC encoder for CTUs that are encoded in the SCF layer. For test set SCID, for example, the average encoding time using the proposed method is 629 seconds, while it is 657 seconds for VVC. This is roughly a gain of 4\%. The symmetric characteristic of the SCF coder leads to an increase in decoding time with an average decoding time of 11 seconds for the proposed method and 0.5 seconds for VVC. However, it should be noted that the proposed method is not optimized for speed, and further optimizations, e.g. regarding the transfer of  data from the VVC decoder to the SCF coder, will further reduce the coding time.															

\section{Summary}
In this paper, we have introduced an improved VVC-based coding method for SCIs using Soft Context Formation. With application of a feature-based block classification technique, parts of an image are encoded using the lossless SCF coder which improves the resulting quality of the image, while, on average, simultaneously reducing the bit rate. We adapt the SCF coder to make use of the information gained by the decoded VVC image by initializing the color palette and the edge pixels for pattern matching and prediction. On average, 4.98\% PSNR-based BD-rate is gained by the proposed method on the evaluated data sets when compared to the unaltered VVC codec. In future work, the segmentation method can be further refined. For example, each block is treated as an independent unit for the proposed segmentation method. However, both in the SCF method and the VVC, causal CTUs will affect the coding efficiency of the current one. In a next step, such dependencies can be investigated and incorporated into the segmentation method for better rate estimation. An enhanced segmentation in combination with a post-processing step may also help in reducing noticeable differences in quality between lossless SCF CTUs and lossy VVC CTUs for small QPs. Finally, a lossy extension of the SCF method could lead to further reduction of bit rate. 
\bibliographystyle{IEEEbib}
\bibliography{refs}

\begin{thebibliography}{10}

\bibitem{Sul12}
{Gary J.} Sullivan, {Jens-Rainer} Ohm, {Woo-Jin} Han, and Thomas Wiegand,
\newblock ``Overview of the high efficiency video coding ({HEVC}) standard,''
\newblock {\em IEEE Transactions on Circuits and Systems for Video Technology},
  vol. 22, no. 12, pp. 1649--1668, Dec. 2012.

\bibitem{Xu16}
Jizheng Xu, Rajan Joshi, and {Robert A.} Cohen,
\newblock ``Overview of the emerging {HEVC} screen content coding extension,''
\newblock {\em {IEEE} Transactions on Circuits and Systems for Video
  Technology}, vol. 26, no. 1, pp. 50--62, Jan. 2016.

\bibitem{Bro21}
Benjamin Bross, Ye-Kui Wang, Yan Ye, Shan Liu, Jianle Chen, Gary~J Sullivan,
  and Jens-Rainer Ohm,
\newblock ``Overview of the versatile video coding ({VVC}) standard and its
  applications,''
\newblock {\em IEEE Transactions on Circuits and Systems for Video Technology},
  vol. 31, no. 10, pp. 3736--3764, Oct. 2021.

\bibitem{VVCSCC}
Tung Nguyen, Xiaozhong Xu, Felix Henry, {Ru-Ling} Liao, {Mohammed Golam}
  Sarwer, Marta Karczewicz, {Yung-Hsuan} Chao, Jizheng Xu, Shan Liu, Detlev
  Marpe, and {Gary J.} Sullivan,
\newblock ``Overview of the screen content support in {VVC}: Applications,
  coding tools, and performance,''
\newblock {\em IEEE Transactions on Circuits and Systems for Video Technology},
  vol. 31, no. 10, pp. 3801--3817, Apr. 2021.

\bibitem{Wei16}
Wei Pu, Marta Karczewicz, Rajan Joshi, Vadim Seregin, Feng Zou, Joel Sole,
  {Yu-Chen} Sun, {Tzu-Der} Chuang, Polin Lai, Shan Liu, {Shih-Ta} Hsiang, Jing
  Ye, and {Yu-Wen} Huang,
\newblock ``Palette mode coding in {HEVC} screen content coding extension,''
\newblock {\em IEEE Journal on Emerging and Selected Topics in Circuits and
  Systems}, vol. 6, no. 4, pp. 420--432, Dec. 2016.

\bibitem{Xu16b}
Xiaozhong Xu, Shan Liu, {Tzu-Der} Chuang, {Yu-Wen} Huang, {Shaw-Min} Lei,
  Krishnakanth Rapaka, Chao Pang, Vadim Seregin, {Ye-Kui} Wang, and Marta
  Karczewicz,
\newblock ``Intra block copy in {HEVC} screen content coding extensions,''
\newblock {\em IEEE Journal on Emerging and Selected Topics in Circuits and
  Systems}, vol. 6, no. 4, pp. 409--419, Dec. 2016.

\bibitem{Xu19}
Xiaozhong Xu, Xiang Li, and Shan Liu,
\newblock ``Intra block copy in versatile video coding with reference sample
  memory reuse,''
\newblock in {\em 2019 Picture Coding Symposium (PCS)}, 2019, pp. 1--5.

\bibitem{Que98}
Ricardo~L. de~Queiroz, Robert~R. Buckley, and Ming Xu,
\newblock ``{Mixed raster content ({MRC}) model for compound image
  compression},''
\newblock in {\em Proc. Visual Communications and Image Processing}, Dec. 1998,
  vol. 3653, pp. 1106 -- 1117.

\bibitem{Kur12}
Ashok~Mathew Kuruvilla,
\newblock ``Tiled image container for web compatible compound image
  compression,''
\newblock in {\em Proc. Eighth International Conference on Signal Image
  Technology and Internet Based Systems}, Nov. 2012, pp. 182--187.

\bibitem{Mog99}
Takeshi Mogi,
\newblock ``A hybrid compression method based on region separation for
  synthetic and natural compound images,''
\newblock in {\em Proc. 1999 International Conference on Image Processing},
  Oct. 1999, vol.~3, pp. 777--781.

\bibitem{Wan14}
Shuhui Wang and Tao Lin,
\newblock ``United coding method for compound image compression,''
\newblock {\em Multimedia Tools and Applications}, vol. 71, no. 3, pp.
  1263--1282, Aug. 2014.

\bibitem{Tan22}
Tong Tang, Ling Li, Xiaoyu Wu, Ruizhi Chen, Haochen Li, Guo Lu, and Limin
  Cheng,
\newblock ``{TSA-SCC}: Text semantic-aware screen content coding with ultra low
  bitrate,''
\newblock {\em IEEE Transactions on Image Processing}, vol. 31, pp. 2463--2477,
  Feb. 2022.

\bibitem{Udd23}
Shabhrish~Reddy Uddehal, Tilo Strutz, Hannah Och, and André Kaup,
\newblock ``Image segmentation for improved lossless screen content
  compression,''
\newblock in {\em Proc. IEEE International Conference on Acoustics, Speech and
  Signal Processing (ICASSP)}, Jun. 2023, pp. 1--5.

\bibitem{Och21}
Hannah Och, Tilo Strutz, and André Kaup,
\newblock ``Optimization of probability distributions for residual coding of
  screen content,''
\newblock in {\em Proc. International Conference on Visual Communications and
  Image Processing (VCIP)}, Dec. 2021, pp. 1--5.

\bibitem{Str19}
Tilo Strutz and Phillip M{\"o}ller,
\newblock ``Screen content compression based on enhanced soft context
  formation,''
\newblock {\em IEEE Transactions on Multimedia}, vol. 22, no. 5, pp.
  1126--1138, May 2020.

\bibitem{JVET_T2013}
Yung-Hsuan Chao, Yu-Chen Sun, Jizheng Xu, and Xiaozhong Xu,
\newblock ``{JVET} common test conditions and software reference configurations
  for non-4:2:0 colour formats,''
\newblock {AHG Report, JVET-T2013}, Joint Video Exploration Team (JVET) of
  ITU-T SG 16 WP 3 and ISO/IEC JTC 1/SC 29/WG 11, {Oct.} 2020.

\bibitem{VVCTSRC}
Tung Nguyen, Benjamin Bross, Heiko Schwarz, Detlev Marpe, and Thomas Wiegand,
\newblock ``Residual coding for transform skip mode in versatile video
  coding,''
\newblock in {\em Proc. 2020 Data Compression Conference (DCC)}, Mar. 2020, pp.
  83--92.

\bibitem{VVCBDPCM}
Mohsen Abdoli, Felix Henry, Patrice Brault, Frederic Dufaux, Pierre Duhamel,
  and Pierrick Philippe,
\newblock ``Intra block-{DPCM} with layer separation of screen content in
  {VVC},''
\newblock in {\em Proc. {IEEE} International Conference on Image Processing
  {(ICIP)}}, Sept. 2019, pp. 3162--3166.

\bibitem{HEVC_SCC_CTC}
Haoping Yu, Robert Cohen, Krishna Rapaka, and Jizheng Xu,
\newblock ``Common test conditions for screen content coding,''
\newblock {JCTVC-U1015-r2}, Joint Collaborative Team on Video Coding (JCT-VC)
  of ITU-T SG 16 WP 3 and ISO/IEC JTC 1/SC 29/WG 11, {Jun.} 2015.

\bibitem{HEVC_RCE3}
Ankur Saxena, Do-Kyoung Kwon, Matteo Naccari, and Chao Pang,
\newblock ``{HEVC} range extensions core experiment 3 ({RCE3}): Intra
  prediction techniques,''
\newblock {JCTVC-N1123}, Joint Collaborative Team on Video Coding (JCT-VC) of
  ITU-T SG 16 WP 3 and ISO/IEC JTC 1/SC 29/WG 11, {Jul.} 2013.

\bibitem{Hua15}
Huan Yang, Yuming Fang, and Weisi Lin,
\newblock ``Perceptual quality assessment of screen content images,''
\newblock {\em IEEE Transactions on Image Processing}, vol. 24, no. 11, pp.
  4408--4421, Nov. 2015.

\bibitem{SCID}
Zhangkai Ni, Lin Ma, Huanqiang Zeng, Ying Fu, Lu~Xing, and {Kai-Kuang} Ma,
\newblock ``{SCID},'' Available:
  \url{http://smartviplab.org/pubilcations/SCID.html}, last accessed: May 28,
  2021 [Online].

\bibitem{Zha17}
Zhangkai Ni, Lin Ma, Huanqiang Zeng, Ying Fu, Lu~Xing, and {Kai-Kuang} Ma,
\newblock ``{SCID}: A database for screen content images quality assessment,''
\newblock in {\em Proc. International Symposium on Intelligent Signal
  Processing and Communication Systems ({ISPACS})}, Nov. 2017, pp. 774--779.

\bibitem{She14}
Chengyao Shen and Qi~Zhao,
\newblock ``Webpage saliency,''
\newblock in {\em Proc. European Conference on Computer Vision}, Sept. 2014,
  pp. 33--46.

\bibitem{GFM}
Zhangkai Ni, Huanqiang Zeng, Lin Ma, Junhui Hou, Jing Chen, and Kai-Kuang Ma,
\newblock ``A gabor feature-based quality assessment model for the screen
  content images,''
\newblock {\em IEEE Transactions on Image Processing}, vol. 27, no. 9, pp.
  4516--4528, May 2018.

\bibitem{Her22}
Christian Herglotz, Matthias Kränzler, Ruben Mons, and André Kaup,
\newblock ``Beyond bjøntegaard: Limits of video compression performance
  comparisons,''
\newblock in {\em Proc. IEEE International Conference on Image Processing
  (ICIP)}, Oct. 2022, pp. 46--50.

\end{thebibliography}

\end{document}